%% file: main.tex
\documentclass[sigconf]{acmart}
\usepackage{algorithmic}
\usepackage{graphicx}
\usepackage{textcomp}
\usepackage{xcolor}
\usepackage{soul}        
\usepackage{array}
\usepackage{multirow}
\usepackage{tabularx}
\usepackage{adjustbox}
\usepackage{framed,enumitem}
\usepackage{hyperref}
\usepackage{array}
\usepackage{float}
\usepackage{booktabs}
\usepackage[utf8]{inputenc}
\usepackage{csquotes}
\newlist{todolist}{itemize}{2}
\setlist[todolist]{label=$\square$}
\usepackage{pifont}
\usepackage[T1]{fontenc}
\usepackage{makecell}
\usepackage{xcolor,pifont}

\newcommand{\bluecheck}{{\color{blue}\ding{52}}}
\newcommand{\redx}{{\color{red}\ding{54}}}

\copyrightyear{2022}
\acmYear{2022}
\setcopyright{acmlicensed}\acmConference[SE4RAI'22 ]{Workshop on Software
Engineering for Responsible AI}{May 19, 2022}{Pittsburgh, PA, USA}
\acmBooktitle{Workshop on Software Engineering for Responsible AI (SE4RAI'22
), May 19, 2022, Pittsburgh, PA, USA}
\acmPrice{15.00}
\acmDOI{10.1145/3526073.3527589}
\acmISBN{978-1-4503-9319-5/22/05}

\begin{document}

\title{Non-Functional Requirements for Machine Learning: An Exploration of System Scope and Interest}


\author{Khan Mohammad Habibullah, Gregory Gay, Jennifer Horkoff}
\email{{khan.mohammad.habibullah, jennifer.horkoff}@gu.se, greg@greggay.com}
\orcid{0000-0001-6794-9585}
\affiliation{%
  \institution{Chalmers | University of Gothenburg}
  \city{Gothenburg}
  \country{Sweden}
}

\renewcommand{\shortauthors}{Habibullah et al.}

\begin{abstract}
Systems that rely on Machine Learning (ML systems) have differing demands on quality---non-functional requirements (NFRs)---compared to traditional systems. NFRs for ML systems may differ in their definition, scope, and importance. Despite the importance of NFRs for ML systems, our understanding of their definitions and scope---and of the extent of existing research---is lacking compared to our understanding in traditional domains.  

Building on an investigation into importance and treatment of ML system NFRs in industry, we make three contributions towards narrowing this gap: (1) we present clusters of ML system NFRs based on shared characteristics, (2) we use Scopus search results---as well as inter-coder reliability on a sample of NFRs---to estimate the number of relevant studies on a subset of the NFRs, and (3), we use our initial reading of titles and abstracts in each sample to define the scope of NFRs over parts of the system (e.g., training data, ML model).  These initial findings form the groundwork for future research in this emerging domain.
\end{abstract}



\maketitle

\section{Introduction}\label{sec:intro}
\input{Introduction}

\section{Background and Related Work}\label{sec:background}
\input{Background}

\section{Methodology}\label{sec:methodology}
\input{Methodology_new}

\section{Results and Discussion}\label{sec:results}
\input{Results}


\subsection{Threats to Validity}\label{sec:threats}
\input{Threats}








\section{Conclusions}\label{sec:conclusion}
\input{Conclusion}

\bibliographystyle{ACM-Reference-Format}
\bibliography{main}
\end{document}

%% file: Introduction.tex
 

Machine Learning (ML) is increasingly being used in decision making and prediction applications that influence many aspects of our lives. Complex systems, referred to as ML systems, use ML to deliver critical functionality. Such systems demand high computational capabilities (often based on GPUs), process a large amount of data, and utilize complex non-deterministic algorithms~\cite{horkoff2019non}. Therefore, ensuring the quality of such systems is potentially more expensive and effort-intensive than traditional systems. 

A thorough requirements engineering (RE) process is necessary to ensure quality. Requirements imposed on system quality are known as non-functional requirements (NFRs), and are expressed over different attributes of quality~\cite{horkoff2019non}. For a ML system, one might imagine constraints over attributes such as fairness, transparency, privacy, security,  or safety~\cite{habibullah2021non}. 

Although significant research has been devoted to NFRs, significant challenges remain for modern system development. Even for traditional systems, NFRs are difficult and challenging to express explicitly~\cite{eckhardt2016non}, and even more difficult to verify or validate~\cite{nuseibeh2000requirements}. Such challenges compound for ML systems, and the identification, definition, and measurement of NFRs for ML systems has emerged as a critical problem to solve~\cite{horkoff2019non,habibullah2021non}.

Much of our accumulated knowledge concerning NFRs may be no longer relevant when dealing with ML systems, due to their complex and non-deterministic nature. Some NFRs, such as fairness, explainability, and privacy become more important. Others, such as usability or interoperability, may become less important~\cite{kamishima2011fairness,obermeyer2019dissecting,habibullah2021non}. New NFRs, such as retrainability, emerge. Moreover, the meaning and interpretation of NFRs for ML systems may differ from traditional systems and may not yet be well understood~\cite{binns2018fairness}. To date, there has been little research on NFRs for ML systems~\cite{horkoff2019non}.
 
 
Further, ``ML'' is not one monolithic entity, but can be considered at different levels of granularity within a larger system~\cite{sculley2015hidden}. When imposing NFRs over an ML system, some NFRs may apply to the algorithm that performs the learning task, while others may apply over the training data used as the basis for such decision making or over the model trained using that data. Still others may apply over the results of applying that model, or over the broader ML system that acts on those results. Therefore, the scope of consideration for NFRs (i.e., the scope of identification, definition, and measurement) for ML systems is a complex and not-yet-solved problem. 

We recently conducted an interview study, which examined treatment of NFRs for ML systems in industry and reported challenges of identifying, defining and measuring NFRs~\cite{habibullah2021non}. Addressing these challenges will require (1) \textbf{a detailed understanding of the definition and scope} of each NFR in a ML system context, and (2), \textbf{an examination of past research} on each of these NFRs as applied to ML system development. These needs are intertwined. To date, there has been no systematic literature reviews or other secondary studies on ML system NFRs. However, performing such a study requires a clear answer to questions of scope to proceed effectively. 

We perform an exploratory study of the treatment of certain NFRs for ML systems in research literature\footnote{While little work has been conducted on the topic of NFRs for ML systems, there is certainly relevant research on individual quality attributes, such as fairness or security.}. Our goals in this study are to (1) gain an approximate idea of the extent to which select NFRs have been studied by researchers, and (2), perform an initial clarification of the scope of these NFRs for ML systems. 

As a starting point, we have taken into account the NFRs identified as important in the interview study~\cite{habibullah2021non}. Using this set of NFRs, we divide NFRs into clusters based on shared attributes of their definitions. This enables understanding of which NFRs could be considered in conjunction. Researchers could study particular clusters, and practitioners may consider defining system quality over related NFRs. We also identify an upper limit on the number of relevant publications in the Scopus database for each NFR. Our initial estimation shows that some NFRs, such as security or transparency, have received significant focus. We select NFRs that have received less attention (e.g., maintainability or testability), and examine the titles and abstracts of 50-100 publications for each. Based on this sample, we estimate the number of relevant publications on each of the selected NFRs. This estimation enables scoping of secondary studies.
Finally, based on inspection of the titles and abstracts of these samples, we perform an exploratory scoping of the selected NFRs in terms of which elements of the system they can be defined over (e.g., training data, ML algorithm, or ML model). This scoping brings further clarity to the specific definitions and treatment of these NFRs, which can benefit future research and practice on each.
 
Our study, while exploratory in nature, is intended to open new opportunities for future research in NFRs for ML systems. We hope to set the groundwork for future studies by clarifying the scoping and definitions for these NFRs, identifying connections between NFRs, and gaining an approximate idea of past interest in these NFRs. Our results can allow researchers to plan future studies and to identify NFRs that have not received sufficient attention. They also help enable engineers to identify which NFRs to consider in conjunction with others of interest, and to think critically about how NFRs apply to different facets of the system-under-development. 



%% file: Background.tex

\noindent\textbf{NFRs for Traditional Systems:} NFRs are considered essential for ensuring the quality of software, but there are no agreed guidelines on how and when NFRs should be elicited, defined, documented, and validated~\cite{glinz2007non}. Moreover, there is no consensus in the requirements engineering (RE) community regarding which step of the RE process NFRs should be considered and applied~\cite{chung2012non}. Significant research has been devoted to NFRs in RE, e.g., Doerr et al. applied a systematic, experience-based, method to elicit, document, and analyze NFRs with the objective of creating a sufficient set of traceable and measurable NFRs~\cite{doerr2005non}. While most work such as this focuses on NFRs for traditional software, we are focused on NFRs specifically for systems that make use of ML to deliver functionality.
Although many researchers have studied NFRs for traditional systems, very few studies to date have focused on NFRs for ML systems.

\smallskip\noindent\textbf{Requirement Engineering for ML Systems:}
Although there are approaches on how to use ML to improve RE tasks, there has not been extensive research on RE for ML systems~\cite{vogelsang2019requirements}. Engineering of ML systems requires different and novel approaches due to their unpredictable nature and differences in their development process. It is crucial to clearly identify and define these differences,  in order to offer tailored practices~\cite{ishikawa2019engineers}. For traditional systems, activities related to requirements analysis and specification are often performed in the early phases of development, with requirements used downstream as part of design, implementation and verification. However, the activity flow often differs for ML systems due to their reliance on data and  the unpredictability of ML results. Upfront problem definition for ML systems can be difficult, as building a clear definition of the problem often requires iterative exploration of data and processes---more so than in typical systems. As such, RE for ML systems has many unknown and unexplored areas, including an understanding of how NFRs differ for such systems.

\smallskip\noindent\textbf{NFRs for ML Systems:}
We discussed challenges and research directions for NFRs for ML systems~\cite{horkoff2019non}. Some of our knowledge about NFRs for traditional systems may no longer be applicable due to the non-deterministic behavior, as well as due to additional performance demands imposed by the need to process and act on large volumes of data. Some NFRs become more important (e.g., explainability), some become less relevant (e.g., modularity), there are differing trade-offs between NFRs (e.g., increased security often causes decreased usefulness), and there is no unified collection and consideration of NFRs for ML-enabled systems. We defined research direction for NFRs for ML systems, including exploring and defining NFRs, as well as reinterpreting and redefining NFRs that already exist for traditional systems. 
 
 
In a recent interview study, we examined challenges regarding NFRs for ML in industry by identifying examples of the identification and measurement of NFRs and examining the importance that practitioners place on NFRs for ML~\cite{habibullah2021non}. The results of the interview study found that most NFRs as defined for traditional software are still relevant for ML-enabled systems. Some NFRs, such as flexibility, efficiency, usability, portability, reusability, and usability were identified as less important by some interviewees. However, they were still considered important by other interviewees. The NFRs identified in the interview study are listed in Table~\ref{Table:ImportantNFRs}. We have defined each NFR, often in an ML context, based on both our experience and related literature, such as research papers, websites, blogs, and forums.  In addition, we reported gaps to address, including identification, definition, scope, and perceptions of NFRs in an ML context~\cite{habibullah2021non}. In this work, we build on these results by beginning to explore the coverage of NFRs in 
research literature. 

\begin{table*}[!t]
    \centering
    \caption{Important NFRs for ML systems, identified in~\cite{habibullah2021non}.} \vspace{-10pt}
    \label{Table:ImportantNFRs}
    \resizebox{\linewidth}{!}{%
\begin{tabular} {|l|l|}
\hline
\textbf{NFRs} &  \textbf{Definition} \\ \hline
Accuracy & The number of correctly predicted data points out of all the data points.\\ \hline
Adaptability & The ability of a system to work well in different but related contexts. \\ \hline
Bias & A phenomenon that occurs when an algorithm produces results that are systematically prejudiced due to erroneous assumptions in the ML process. \\ \hline
Completeness & An indication of the comprehensiveness of available data, as a proportion of the entire data set, to address specific information requirements. \\ \hline
Complexity & When a system or solution has many components, interrelations or interactions, and is difficult to understand. \\ \hline
Consistency & A series of measurements of the same project carried out by different raters using the same method should produce similar results. \\ \hline
Correctness & The output of the system matches the expectations outlined in the requirements, and the system operates without failure. \\  \hline
Domain Adaptation & The ability of a model trained on a source domain to be used in a different---but related---domain.  \\ \hline
Efficiency & The ability to accomplish something with minimal time and effort. \\ \hline
Ethics & Concerned with adding or ensuring moral behaviors. \\ \hline
Explainability & The extent to which the internal mechanics of ML-enabled system can be explained in human terms. \\ \hline
Fairness & The ability of a system to operate in a fair and unbiased manner \\ \hline
Fault Tolerance & The ability of a system to continue operating without interruption when one or more of its components fail. \\ \hline
Flexibility & The ability of a system to react to changing  demands or conditions.\\ \hline
Integrity & The ability to ensure that data is real, accurate and safeguarded from unauthorised modification. \\ \hline
 Interpretability & The extraction of relevant knowledge from a model concerning relationships either contained in data or learned by the model \\ \hline
Interoperability & The ability for two systems to communicate effectively \\ \hline
Justifiability & The ability to be show the output of an ML-enabled system to be right or reasonable. \\ \hline
Maintainability & The ease with which a system or component can be modified to correct faults, improve performance or other attributes, or adapt to a changed environment \\ \hline
Performance & The ability of a system to perform actions within defined time or throughput bounds. \\ \hline
Portability & The ability to transfer a system or element of a system from one environment to another. \\ \hline
Privacy & An algorithm is private if an observer examining the output is not able to determine whether a specific individual's information was used in the computation. \\ \hline
Reliability & The probability of the software performing without failure for a specific number of uses or amount of time. \\ \hline
Repeatability & The variation in measurements taken by a single instrument or person under the same conditions. \\ \hline
Retrainability & The ability to re-run the process that generated the previously selected model on a new training set of data. \\ \hline
Reproducibility & One can repeatedly run your algorithm on certain datasets and obtain the same (or similar) results. \\ \hline
Reusability &  The ability of reusing the whole or the greater part of the system component for similar but different purpose.  \\ \hline
Safety &  The absence of failures or conditions that render a system dangerous \\ \hline
Scalability & The ability to increase or decrease the capacity of the system in response to  changing demands. \\ \hline
Security & Security measures ensure a system's safety against espionage or sabotage. \\ \hline
Testability &  The ability of the system to to support testing by offering relevant information or ensuring the visibility of failures. \\ \hline
Transparency & The extent to which a human user can infer why the system made a particular decision or produced a particular externally-visible behaviour. \\ \hline
Traceability & The ability to trace work items across the development lifecycle. \\ \hline
Trust & A trusted system is a system that is relied upon to a specified extent to enforce a specified security, or a security policy. \\ \hline
Usability & How effectively users can learn and use a system. \\ \hline
\end{tabular}}
\vspace{-10pt}
\end{table*}



 

\smallskip\noindent\textbf{ML as Part of a Larger Software System:}
In a ML system, the ``ML'' is a small part of a larger system~\cite{sculley2015hidden}. In traditional systems, NFRs can be identified over the whole system or elements of the system. In an ML context, NFRs can also be defined over different parts of the system. These elements may differ from traditional systems, and the differing nature of these elements may lead to a differing understanding of relevant NFRs. In our preliminary NFR definitions in Table~\ref{Table:ImportantNFRs}, we have sometimes defined NFRs in ML terms, referencing the ML model or data.  However, this is not done consistently, and not all potential elements of the system are considered. In an effort to improve how NFRs for ML systems are defined, we explore the idea of NFR ``scope'' further in this study.


%% file: Methodology_new.tex
Though NFRs for traditional software are fairly well-understood, there are still gaps in our foundational knowledge on NFRs for ML systems. We are eager to learn which NFRs for ML systems have been explored by other researchers and which are yet to be investigated heavily. We also want to learn how 
NFRs for ML are perceived by other researchers so that the definitions and scopes of such NFRs can be refined. 

Hence, we have performed an exploratory study aimed at at establishing an initial scoping of the treatment of NFRs for ML and an initial estimation of the level of research that has been conducted on these NFRs.
A systematic mapping study is primarily concerned with structuring a research area~\cite{kitchenham2007guidelines}. As we are performing an initial exploration of the scope of NFRs for ML systems, we have adapted the systematic mapping study concept for our purposes.

Our goals in this study are to (1) gain an approximate idea of the extent to which select NFRs have been studied by researchers, and (2), perform an initial clarification of the scope of these NFRs for ML systems. 
Specifically, we address the following research questions:

\smallskip\noindent \textbf{RQ1:} Can the ML system NFRs be grouped into a small number of clusters based on shared characteristics?

\noindent \textbf{RQ2:} Which NFRs have received the most---or least---attention in existing research literature?

\noindent \textbf{RQ3:} Over which elements of an ML system can NFRs be defined?

\smallskip\noindent To answer these questions, we grouped the NFRs into a small number of clusters based on their shared characteristics (Sec.~\ref{sec:Clustering}). We performed the initial stages of a mapping study in order to gain a rough approximation of the how much research exists on each NFR---focusing on those NFRs that have been least investigated or belong to two particular clusters of interest (Sec.~\ref{sec:Publication_Volume_Estimation}).
Then, based on the titles and abstracts and past experience, we identify which elements of the system that these NFRs should be defined and measured over (Section~\ref{sec:NFRScopeDetermination}).

\subsection{NFR Clustering}\label{sec:Clustering}

In Table~\ref{Table:ImportantNFRs}, we listed the NFRs found to be important in the interview study~\cite{habibullah2021non}. For each, we have defined them based both on our past experience and based on their treatment in a small sampling of research papers, websites, blogs, and forums. Based on these definitions, we are interested in grouping these NFRs into a small number of clusters, where each cluster contains NFRs that have similar meaning or purpose. Researchers can use these clusters to identify which NFRs may be related and able to collectively determine the quality of a system. Researchers could also perform secondary studies on particular clusters of NFRs. Developers can also use these clusters to identify which NFRs may be relevant to their particular needs or system-under-development.

We have created these clusters primarily through discussion of the NFRs and their definitions. During a series of meetings, we read and interpreted the definitions and debated their meaning. We then discussed and decided which cluster to assign an NFR to. We have placed NFRs in clusters if they are a similar purpose within system development or could be measured in a similar manner. 

For example, the explainability of a ML system refers to the extent to which its internal mechanics can be explained in human terms. Transparency refers to the ability of the system to clarify the reasoning for its decisions to a human user. These NFRs differ in their exact meaning and assessment, but are both key elements in ensuring that ML systems operate in a clear and reasonable manner. Therefore, both should reside in the same cluster.

We also created a separate cluster for those NFRs that could not be put into any of the other clusters, as they lacked any shared characteristics with the NFRs in other clusters.

Our goal at this stage is not to create a formal hierarchy, as exists for NFRs for traditional systems~\cite{boehm1976quantitative}. Rather, our interest is in creating a lightweight organizational structure for use in understanding the scoping and definition of NFRs for ML systems. 
 
\subsection{Publication Volume Estimation} \label{sec:Publication_Volume_Estimation}

In this section, we describe our strategy for estimating the number of research papers for certain NFRs.

\vspace{-0.1cm}
\subsubsection{Initial Paper Search}\label{sec:initial_search}

We performed a database search---including all publications up to September 2021---in order to identify the research papers that may be relevant for each NFR. We selected Scopus, a meta-database, which includes research papers from peer-reviewed journals and conferences from multiple publishers such as IEEE, ACM, and Elsevier. Scopus is considered one of the most representative and rich in content for Software Engineering research and is used in many secondary studies~\cite{keele2007guidelines}.

We identified relevant search terms and developed search strings for the database search. We first derived the major terms (e.g., machine learning, non-functional requirements). Then, we identified synonyms or alternative spelling for the major terms from related literature, and based on our discussions. We also split major terms into more specific and clear terms. For example, we split the general term ``non-functional requirements'' into strings based on specific NFRs. Finally, we concatenated these terms to form search strings. 


We apply one search string per NFR. The string includes that NFR, as well as terms related to machine learning: \textbf{(``machine learning'' OR ``supervised learning'' OR ``unsupervised learning'' OR ``reinforcement learning'' OR ``deep learning'')}. 

For example, to identify papers on interoperability, we have used the search string: \textbf{(``machine learning'' OR ``supervised learning'' OR ``unsupervised learning'' OR ``reinforcement learning'' OR ``deep learning'') AND (``interoperability'')}. As a second example, to identify papers related to usability, we have used the string: 
 \textbf{(``machine learning'' OR ``supervised learning'' OR ``unsupervised learning'' OR ``reinforcement learning'' OR ``deep learning'') AND (``usability'')}.
 
The number of papers found from this step give an upper limit on the number of relevant publications. Not all of these publications are likely to be relevant, as they may not relate to the use of such properties as NFRs for a ML system. For example, 
several of the results for maintainability described work which used ML to predict maintainability of another system, rather than focusing on maintainability of an ML system.  
Therefore, in the next step, we used a sample of publications to gain a finer estimation of the number of relevant publications for a subset of the NFRs.

\subsubsection{NFR Selection}\label{sec:NFR_selection}

This upper limit gives some indication of the research interest in each NFR. To gain a clearer estimation of the percentage of those publications that are relevant, we have chosen to focus on a subset of the list of NFRs. Some NFRs, such as performance or security, have already received significant attention from the research community. We would recommend that future secondary studies focus specifically on these topics.  We have instead focused on those NFRs that have received less focus from researchers, including those with a lower number of publications as well as those that we identified as being part of two clusters of interest (the ``other'' cluster and a cluster centered around tailoring a system to different environments).  

We created a list of the number of publications found in the search results for each NFR. At first, we sorted the NFRs based on the number of publications, in decreasing order. We then excluded those NFRs that have more than 1,200 search results. For example, we excluded accuracy, as the number of retrieved papers was more than the threshold. Based on this threshold, we excluded 16 NFRs. 

We then took into account which cluster we assigned each NFR to. If an NFR has more search results than the threshold but falls into the two clusters that we selected for initial inspection, then we included that NFR for consideration. As a result, we reincorporated usability and flexibility into our estimation, as those NFRs fall into these two clusters even though those have more search results than the threshold. We perform a more detailed analysis on 20 NFRs. 

\begin{figure*}[!t]
    \centering
    \includegraphics[width=0.9\textwidth]{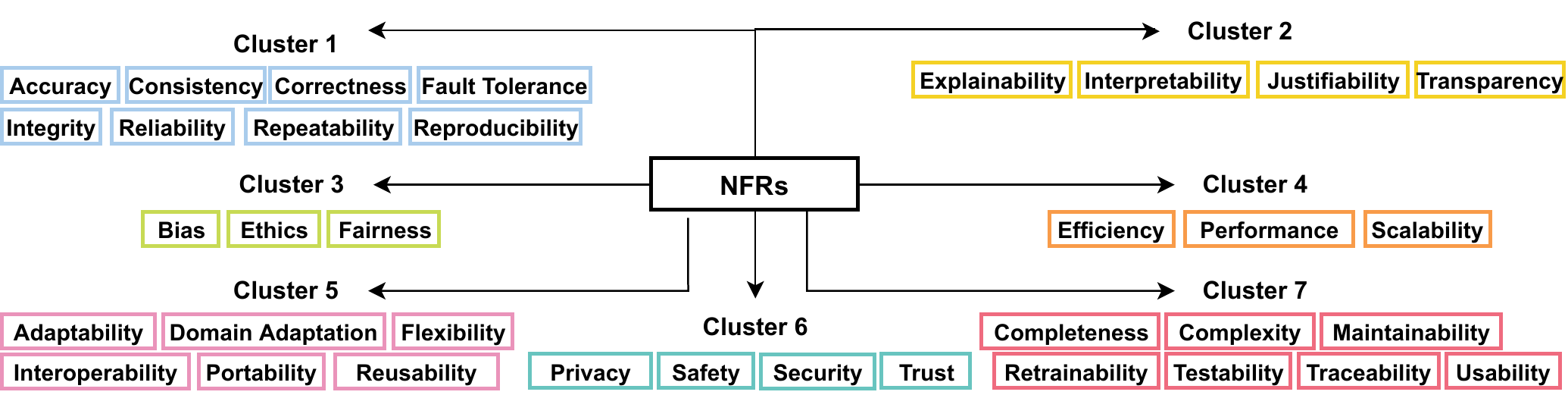} \vspace{-10pt}
    \caption{NFRs divided into clusters, based on shared characteristics.}
    \label{fig:NFRsClusters} \vspace{-10pt}
\end{figure*} 

\subsubsection{Estimating the Number of Relevant Papers for Selected NFRs} \label{sec:Estimation_refinement}

We estimate the number of relevant publications for each selected NFR by inspecting the titles and abstracts of a sample of 50 papers. We read the title and abstract of each publication and use inclusion and exclusion criteria to filter these publications, marking them as relevant or irrelevant. Each author determined the relevancy of each paper independently. We then discussed each disagreement in a meeting, using our criteria, and formed a final list. 

\smallskip\noindent\textbf{Inclusion Criteria:} The publication must discuss an NFR from Table~\ref{Table:ImportantNFRs}. It must focus on the definition, identification, measurement, or challenges of a NFR for a ML system, or for an element of the system (e.g., the model). It must have been published in a peer-reviewed journal, conference, or workshop. The full text must be accessible and written in English.

\smallskip\noindent\textbf{Exclusion Criteria:} The publication is focused on topics other than NFRs for ML systems. This includes publications where ML is used to measure, improve, or predict a NFR. For example, the authors used ML to classify requirements into different NFRs~\cite{kurtanovic2017automatically}. In such a case, the publication is not relevant for examining how such an NFR affects the development of a ML system. The publication simply uses the NFR as an evaluation criteria, but does not discuss or describe the use of the NFR during system development. For example, if an author uses completeness as part of their evaluation of the results of a system, but the actual research has no relation to improving the completeness of a ML system, then it is excluded. The publication was not written in English, not peer-reviewed, or lacks an available full text. Editorials, abstracts, book chapters, workshop summaries, poster sessions, prefaces, article summaries, interviews, news, reviews, comments, news, reviews, tutorials, panels, and discussions are excluded.

\smallskip\noindent\textbf{Inter-coder Reliability:} 
Following the process of individually reviewing 50 papers for selected NFRs, we calculated our agreement using Fleiss' kappa, a statistical measure for assessing ICR between a fixed number of raters. 
In some cases where the ICR was low, or where there were significant disagreements, we repeated the sampling process for a second set of 50 papers. In such cases, it was hoped that we could clarify our shared definition and estimation of the scope of the NFR. If the ICR either increased or stayed the same, this served to increase confidence in our understanding.



The final list of relevant papers, after discussion, gives an indication of the number of publications that may be relevant from that initial set retrieved from Scopus. This, in turn, offers an indication of research interest in the NFR. 


\smallskip\noindent\textbf{Estimating the Number of Publications:} 
We counted the number of publications that were deemed relevant from the first---and, in some cases, second---sample for each selected NFR. We used these counts along with the total number of papers found by Scopus to estimate the total number of included papers. This estimation is calculated by simply multiplying the total number of publications by the percentage of the sample that was deemed relevant. 

For example, we found an upper limit of 851 publications for the transparency NFR. After screening 50 publications, we agreed to include 44 (88\%). Extending to the full set of 851 papers, we estimate that 749 publications will actually be relevant. As a second example, we identified 214 publications for traceability. In this case, we sampled 100 publications, and decided that 10 were relevant (10\%). Therefore, approximately 21 of the 214 are expected to be relevant to the treatment of the property as a NFR for ML systems.

We repeated this calculation for the rest of the selected NFRs, producing an estimation of the number of relevant papers for each. This is still a rough approximation of past research interest, but it is sufficient to provide an initial portrait of the field and to refine our own definitions and ideas regarding scope.

 

 

\subsection{NFR Scope Determination}\label{sec:NFRScopeDetermination}

In order to clearly define or measure the attainment of an NFR, it must be understood exactly how the NFR applies to the system. This determination requires understanding whether a NFR relates to the system as a whole, or perhaps to a lower level of granularity within the system. In the case of a ML system, an NFR may be defined and measured over different aspects of the ML application. For example, an NFR may apply differently when we discuss the training data, the algorithm that uses the training data to build a model, or to the model trained on that data. 

Therefore, we have first determined which elements of a ML system are particularly relevant when we discuss the NFRs for such systems. 
We then used our existing definitions, past experience, and the titles and abstracts of the relevant studies examined in the previous step in order to determine to which of these elements  each NFR was applicable. In a series of meetings, we discussed each NFR in relation to these system elements. In each case, we made a determination by coming to an agreement and discussing any cases where we disagreed---generally by identifying an example of how that NFR is applied to that element.  For example, repeatability refers to the level of variability in the behavior of the system. Repeatibility is a property of the results---or of the system as a whole---rather then a property of the model, algorithm, or training data. It is the results that vary, not the model itself. 
This scoping is intended as a starting point for establishing detailed definitions for each NFR in an ML system context.

%% file: Results.tex


\smallskip\noindent\textbf{NFR Clustering (RQ1):}
We were able to create six different clusters, where each cluster includes the NFRs that share similar properties and purposes. For example, after analyzing their definitions, we found that ethics, bias, and fairness shares similar meanings and serve similar purposes. Therefore, we put these three NFRs into the same cluster. 
 
The clusters are presented in Fig.~\ref{fig:NFRsClusters}. Cluster 1 includes NFRs that are related to assessing the functional correctness of ML systems and aspects of correctness. This includes the core correctness, as well as assessment of correctness (e.g., accuracy) and variance (e.g., reliability, consistency). Cluster 2 contains NFRs related to understanding the internal decisions or results of applying ML (e.g., transparency, explainability). The NFRs related to ethical aspects of ML systems, such as fairness and bias, form cluster 3. NFRs related to the performance (e.g., speed) of an ML system are contained in cluster 4. The qualities related to tailoring and adjustment of the ML system to different environments (e.g., flexibility, adaptability) are grouped in cluster 5. Concerns related to privacy and security are grouped together in cluster 6. The NFRs that do not share similar properties are grouped in cluster 7.

Previous work has presented NFRs in terms of a hierarchy (e.g.,~\cite{cavano1978framework}) or as part of a interdependency graph (e.g.,~\cite{chung2012non}). Our goal was not to suggest a definite hierarchy, but to group NFRs to clarify relatedness and scope, particularly for future research studies. For example, a study may focus on a particular cluster or one or two related NFRs. These clusters can also help practitioners  understand the similarity of NFRs and provide guidance on which related NFRs they should consider while developing ML systems.
 
 
\begin{table}[!t]
  \centering
  \scriptsize
  \caption{NFRs with number of search results, number of relevant publications, kappa values (agreement on sample), and final paper volume estimation for select NFRs. We only examined a second sample in cases where we wanted to see if agreement would improve.} \label{tab:metadata} \vspace{-5pt}
    \begin{tabular}{|l|c|c|c|c|c|c|}
    \hline
    \textbf{NFR} & \textbf{Search}  & \textbf{Relevant} & \textbf{Kappa} & \textbf{Relevant} & \textbf{Kappa} & \textbf{Est.} \\ 
    & \textbf{Results} & \textbf{(1)} & \textbf{(1)} & \textbf{(2)} & \textbf{(2)} & \textbf{Pubs.}\\
   \hline
   Performance & 114853  & \multicolumn{5}{c}{}  \\ \cline{1-2}
    Accuracy & 92669  &  \multicolumn{5}{c}{}  \\
    \cline{1-2}
    Efficiency & 22247  &  \multicolumn{5}{c}{}  \\
    \cline{1-2}
    Security & 19142  &   \multicolumn{5}{c}{}  \\
    \cline{1-2}
    Complexity & 16997 & \multicolumn{5}{c}{}  \\
    \cline{1-2}
    Privacy & 6388   &   \multicolumn{5}{c}{}  \\
    \cline{1-2}
    Safety & 5848   &     \multicolumn{5}{c}{}  \\
    \cline{1-2}
    Reliability & 5620   &   \multicolumn{5}{c}{}  \\
    \cline{1-2}
    Bias  & 4118   &     \multicolumn{5}{c}{}  \\
    \cline{1-2}
    Scalability & 3595   &   \multicolumn{5}{c}{}  \\
    \cline{1-2}
    Consistency & 2936   &   \multicolumn{5}{c}{}  \\
    \cline{1-4}\cline{7-7}
    Flexibility & 2764   & 23 (46\%) & 0.54  &   \multicolumn{2}{c|}{}  & 1271 \\
    \cline{1-4}\cline{7-7}
    Interpretability & 2418   & \multicolumn{5}{c}{}  \\
    \cline{1-2}
    Trust & 1965   &  \multicolumn{5}{c}{}  \\
    \cline{1-2}
    Reproducibility & 1796   &  \multicolumn{5}{c}{}  \\
     \cline{1-4}\cline{7-7}
    Domain Adapt. & 1732   & 47 (94\%) & 0.63  &  \multicolumn{2}{c|}{}    & 1628 \\
    \hline
    Usability & 1270  & 21 (42\%) & 0.50 & 29 (58\%) & 0.44 & 635 \\
   \hline
    Adaptability & 1177   & 34 (68\%) & 0.50  &       \multicolumn{2}{c|}{}     & 800 \\
    \cline{1-4}\cline{7-7}
    Fairness & 1089   & 45 (90\%) & 0.41  &  \multicolumn{2}{c|}{}       & 980 \\
    \cline{1-4}\cline{7-7}
    Correctness & 1045   & 16 (32\%) & 0.53  &  \multicolumn{2}{c|}{}    & 334 \\
    \cline{1-4}\cline{7-7}
    Integrity & 1015   &    \multicolumn{5}{c}{} \\
    \cline{1-4}\cline{7-7}
    Transparency & 851    & 44 (88\%) & 0.70  &       \multicolumn{2}{c|}{}     & 749 \\
    \cline{1-4}\cline{7-7}
    Explainability & 706    & 44 (88\%) & 0.22  &      \multicolumn{2}{c|}{}    & 621 \\
    \cline{1-4}\cline{7-7}
    Fault Tolerance & 553    & 26 (52\%) & 0.68  &    \multicolumn{2}{c|}{}    & 288 \\
   \cline{1-4}\cline{7-7}
    Interoperability & 532    & 9 (18\%) & 0.45  &    \multicolumn{2}{c|}{}   & 96 \\
    \hline
    Completeness & 372  & 23 (46\%) & 0.40 & 25 (50\%) & 0.58 & 179 \\
   \hline
    Portability & 346    & 21 (42\%) & 0.45  &   \multicolumn{2}{c|}{}        & 145 \\
   \cline{1-4}\cline{7-7}
    Ethics & 331    & 31 (62\%) & -0.03 &  \multicolumn{2}{c|}{}     & 205 \\
   \cline{1-4}\cline{7-7}
    Reusability & 321   & 24 (48\%) & 0.55  &  \multicolumn{2}{c|}{}        & 154 \\
   \hline
    Maintainability & 277  & 6 (12\%) & 0.30 & 9 (18\%) & 0.72 & 42 \\
    \hline
    Traceability & 214  & 4 (8\%) & 0.61 & 6 (12\%) & 0.61 & 21 \\
    \hline
    Repeatability  & 171   & 17 (34\%) & 0.44  &       \multicolumn{2}{c|}{}    & 58 \\
    \hline
    Testability & 77  &4 (8\%) & 0.54 & 2 (4\%) & 1.00 & 5 \\
    \hline
    Justifiability  & 3     & 0 (0\%) & 1.00  &       \multicolumn{2}{c|}{}     & 0 \\
   \cline{1-4}\cline{7-7}
    Retrainability & 0  &  \multicolumn{4}{c|}{} & 0\\
    \cline{1-2}\cline{7-7}
    \end{tabular} \vspace{-10pt}
\end{table}
 
\smallskip\noindent\textbf{Estimated Number of Publications (RQ2):}
We used the search strings described in Sec.~\ref{sec:Publication_Volume_Estimation} to identify an upper limit on the number of relevant publications for each NFR. The number of identified publications is presented the second column of Table~\ref{tab:metadata}. We found the most results for performance, accuracy, and efficiency; while, repeatability, testability, and justifiability yielded the fewest results. We found no research papers in Scopus for retrainability, potentially indicating that this term is not common. 

 

 We can sum the total number of search results for each cluster, finding that cluster 4 (performance, \ldots) has 140695 results, cluster 1 (accuracy, \ldots) 105805, cluster 6 (security, \ldots) 33343, cluster 7 (``other'' NFRs) 19207, cluster 5 (adaptability, \ldots) 6872, cluster 3 (bias, \ldots) has 5538 and cluster 2 (explainability, \ldots) has 3978 results. 
 
We can reflect on the number of papers found via the Scopus search. The number of papers for accuracy is very high, as  researchers and practitioners are very focused on prediction accuracy. We also found more papers for usability than we expected, even when excluding papers using usability as a synonym for applicability, and find it encouraging that research is focusing on human-oriented aspects.  

We were surprised that no publications were found for retrainability, even though practitioners mentioned retrainability as important~\cite{habibullah2021non}. We hypothesize that these ideas are being discussed using alternative terms. Similarly, we were surprised by few search results on testability, but this may again be due to use of different terms. We also expected more search results on fairness, as we perceive that researchers and practitioners are focusing more on this topic. This may be due to commonality and split of results amongst bias, fairness, and ethics. 

The performance and accuracy clusters (4 and 1) show the most raw results, followed by the security cluster. These results are generally in line with our expectations. We can see a particular interest in cluster 4, including performance.  In ML terminology, performance often denotes a form of accuracy or correctness, as opposed to time or resource usage---as the term is often used in typical SE. Cluster 7 also has a relatively large number of results, mainly due to the inclusion of complexity.  

We were surprised that clusters 2---containing explainability---and cluster 3---containing bias---yielded relatively fewer results. Even though these are perceived as hot topics in research, either the volume of papers is still relatively small, this work includes terms which differ from the NFRs included as part of our search. 

Although these results are a useful staring point, we refine our estimation for a subset of NFRs to estimate how many publications are relevant. When selecting NFRs for a more detailed estimation, we focused on NFRs that are less researched (but still potentially important), and those in Clusters 5 and 7. 
 We applied the inclusion-exclusion criteria and ICR process described in Sec.~\ref{sec:Estimation_refinement} for a sample of fifty papers of each selected NFR. We present the number of publications found to be relevant for each, along with the inter-coder reliability in Table~\ref{tab:metadata}. In some cases, we also conducted a second sample of an additional 50 publications. 
 
 
We can evaluate the strength of our agreement as follows: $<$ 0.0 is considered as poor agreement, 0.00-0.20 as slight, 0.21-0.40 as fair, 0.41-0.60 as moderate, 0.61-0.80 as substantial, and 0.81-1.00 as almost perfect~\cite{landis1977measurement}. We attained a substantial rage of scores in terms of ICR for the NFRs. One result (ethics) was poor, with our coding being worse than random. However, we attained fair results for three other NFRs, and moderate or better for the remaining 15. Our final estimation is shown in the final column in Table.~\ref{tab:metadata}.  
 
Focusing on five of the NFRs in cluster 7, we can examine our change in agreement after discussion. After the first sample of 50 papers, the ICR scores were fair for maintainability and completeness, and moderate for usability, completeness, maintainability, traceability, and testability. After a discussion among all three authors about our perception and interpretation of the NFR definitions and the inclusion and exclusion criteria, the ICR for the second sample generally improved and ranged between moderate (e.g., completeness, maintainability, traceability) to perfect (e.g., testability). We note, however, that our ICR score for usability actually decreased in the second round.  To some extent, these score also depend on the percentage of relevant publications. The less often papers are relevant (e.g., testability), the easier it is to gain high agreement.  

For some NFRs, it was difficult to agree on inclusion. For example, we had particularly low agreement for ethics and explainability. For some NFRs, we can often make a clear distinction between studies focused on improving attainment of that NFR when designing a ML system versus irrelevant studies (e.g., those that use ML to predict attainment of an NFR for a traditional system). With topics like ethics and explaininability, it was harder to make this type of distinction, and there were more disagreements on particular studies. In these cases, future work may require clearer definitions or more specific criteria. 
 

\smallskip\noindent\textbf{NFR Scoping Over System Elements (RQ3):}
Clear definition of NFRs in a ML system context,requires understanding which specific elements of an ML system that an NFR is applicable to. 

As a starting point for building this understanding, we believe that NFRs can be defined over the following parts of an ML system. The \textbf{training data} used by the ML algorithm as the basis for making decisions. The \textbf{algorithm} that performs the learning task. This includes algorithms that operate on training data, as well as those that perform learning tasks based on feedback, such as reinforcement learning agents. We also consider the specific implementation of the algorithm here.The core \textbf{model or artifact\footnote{This notion also encompasses the policy learned by an agent in reinforcement learning, or other rules ``learned'' by the algorithm in other techniques.}} built by the algorithm for use in making decisions. For example, the algorithm may use the training data to build a model that makes decisions in new situations based on learned connections between data items. The \textbf{resulting decisions or behaviors} made as a result of applying the model. Finally, \textbf{the ML system as a whole}.

\begin{figure}[!t]
    \centering
    \includegraphics[width=\linewidth]{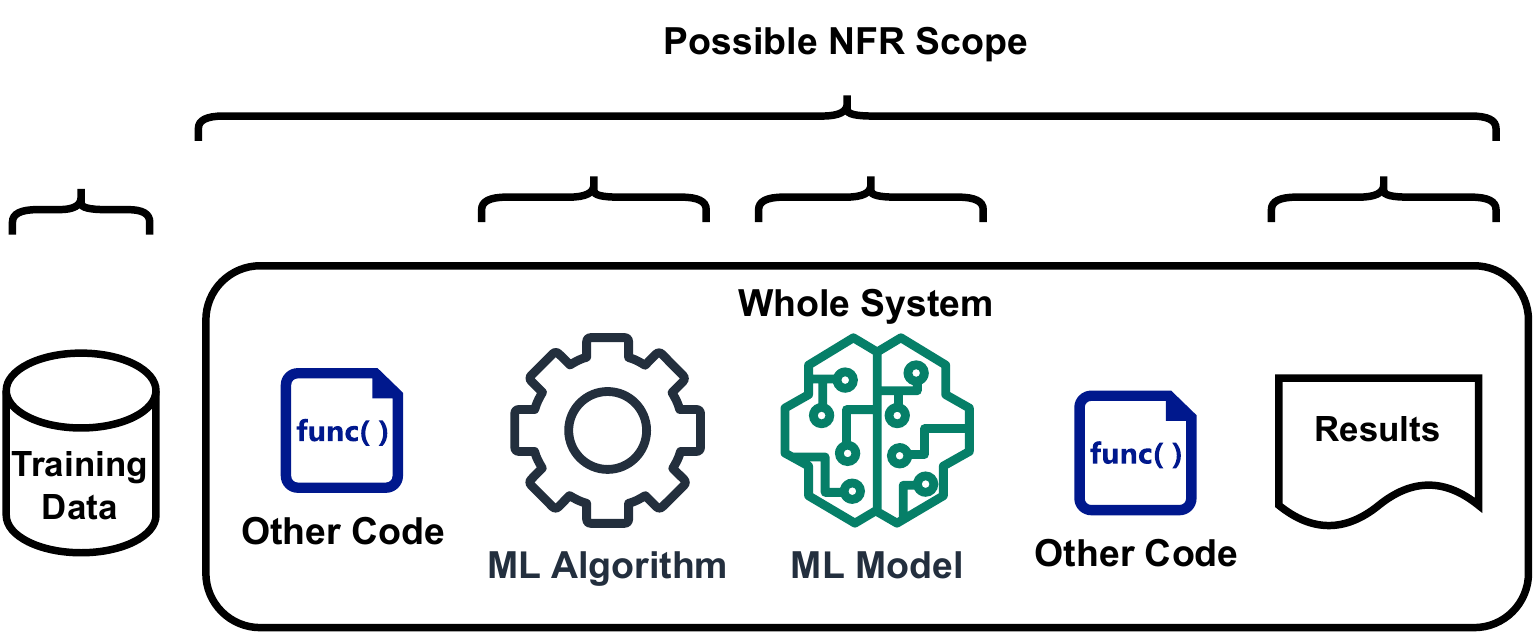} \vspace{-5pt}
    \caption{Possible scope for NFRs over system elements.}
    \label{fig:MLsmall} \vspace{-10pt}
\end{figure}

These parts are illustrated in Fig.~\ref{fig:MLsmall}. 
It is possible that more elements may be applicable in the future, e.g., NFRs over features of a data set or over specific types of functionality operating on the results of ML, but we start with this initial list of system elements to understand the scope of the selected NFRs.  
 
\begin{table}[!b]
  \centering
  \scriptsize
  \caption{System elements that NFRs can be defined over}   \label{tab:nfrdefinedover} \vspace{-5pt}
    \begin{tabular}{|p{8.35em}|c|c|c|c|c|c|}
    \cline{3-7}
    \multicolumn{2}{c}{} & \multicolumn{5}{|c|}{\textbf{System Element the NFR Can be Defined Over}} \\
    \hline
    \textbf{NFR} & \textbf{Cluster} & \textbf{Train.} & \textbf{Algo.} & \textbf{Model} & \textbf{Results} & \textbf{Whole} \\
    & & \textbf{Data} & & & & \textbf{System} \\
    \Xhline{3\arrayrulewidth}
    Completeness & 1 & \bluecheck   & \redx    & \bluecheck   & \redx    & \bluecheck \\
    \hline
    Correctness & 1 & \bluecheck   & \bluecheck   & \bluecheck   & \bluecheck   & \bluecheck \\
    \hline
    Fault Tolerance & 1 & \redx    & \bluecheck   & \bluecheck   & \redx    & \bluecheck \\
    \hline
    Integrity & 1 & \bluecheck   & \bluecheck   & \bluecheck   & \bluecheck   & \bluecheck \\
    \hline
    Repeatability & 1 & \redx    & \redx    & \redx    & \bluecheck   & \bluecheck \\
    \Xhline{3\arrayrulewidth}
    Explainability & 2 & \redx    & \bluecheck    & \bluecheck   & \bluecheck   & \bluecheck \\
    \hline
    Transparency & 2 & \redx    & \bluecheck   & \bluecheck   & \bluecheck   & \bluecheck \\
    \Xhline{3\arrayrulewidth}
    Ethics & 3 & \bluecheck   & \bluecheck   & \bluecheck   & \bluecheck   & \bluecheck \\
    \hline
    Fairness & 3 & \bluecheck   & \bluecheck   & \bluecheck   & \bluecheck   & \bluecheck \\
    \Xhline{3\arrayrulewidth}
    Adaptability & 5 & \bluecheck   & \bluecheck   & \bluecheck   & \bluecheck   & \bluecheck \\
    \hline
    Domain Adaptation & 5 & \bluecheck   & \bluecheck   & \bluecheck   & \bluecheck   & \bluecheck \\
    \hline
    Flexibility & 5 & \redx & \bluecheck   & \bluecheck   & \redx    & \bluecheck \\
    \hline
    Interoperability & 5 &  \redx    & \bluecheck   & \bluecheck   & \redx    & \bluecheck \\
    \hline
    Portability & 5 & \bluecheck   & \bluecheck   & \bluecheck   & \redx    & \bluecheck \\
    \hline
    Reusability & 5 & \bluecheck   & \bluecheck   & \bluecheck   & \redx    & \bluecheck \\
    \Xhline{3\arrayrulewidth}
    Maintainability & 7 & \bluecheck   & \bluecheck   & \bluecheck   & \redx    & \bluecheck \\
    \hline
    Testability & 7 & \redx    & \bluecheck   & \bluecheck   & \bluecheck    & \bluecheck \\
    \hline
    Traceability & 7 & \bluecheck   & \bluecheck   & \bluecheck   & \bluecheck   & \bluecheck \\
    \hline
    Usability & 7 & \redx    & \bluecheck   & \bluecheck   & \bluecheck   & \bluecheck \\
    \hline
    \end{tabular}%
\vspace{-10pt}
\end{table}%
 
Our overall determination of which system elements a particular NFR can be defined over is presented in Table.~\ref{tab:nfrdefinedover}. 
Note that our estimation of NFR scope is an initial estimation based on our experiences and the sampled abstracts. The scope of each NFR likely will evolve over time as more data and examples are gathered.

To illustrate our determinations, we select a number of examples. For example, we determined that the NFR flexibility can be defined over the ML algorithm, the ML model, and the whole system. However, we believe it is not applicable to the training data and the results. Consider a definition of flexibility by Ladiges et al.~\cite{ladiges2013operationalized}, ``\textit{flexibility is an indicator for the ability of a system to react to changing  demands or conditions}''. We can adapt this definition to different parts of the ML system, as in the following\footnote{We note that these definitions may have  significant overlap with definitions for NFRs such as adaptability, resuability, or portability, which is precisely why these NFRs are placed in the same cluster---cluster 5, in this case.}.
The flexibility of an ML algorithm is \textit{the ability of an algorithm to react to changing demands and conditions, without significant re-implementation.} The flexibility of an ML model is \textit{the ability of a model to react to changing inputs and contexts in a useful way, without retraining.} Finally, the flexibility an ML system could use the initial definition or, more specifically, \textit{the ability of a ML system to react to changing demands or conditions without extensive re-implementation or re-training.} On the other hand, we struggle to define flexibility over training data. It makes sense to think of the reusability of training data, e.g., to train ML systems for different context and purposes with some of the same data, but what does it mean for data itself to be flexible? Similarly, results can be reusable, but it is not clear how they can be flexible.  We opt to omit these definitions from our consideration.
 

Similarly, the NFR usability can be defined over the ML algorithm, the ML model, the results, and the whole system; but may not be applicable over the training data. If we take the simple definition of usability from Table~\ref{Table:ImportantNFRs}, ``\textit{how effectively users can learn and use a system}'', this definition makes sense over the whole system. We can also define this NFR over specific ML elements. The usability of an ML algorithm is \textit{how effectively users can learn and use an algorithm to train an ML model as part of a system.} The usability of an ML model is \textit{how effectively users learn to use an ML model at run-time in order to get results}. The usability of ML results is \textit{how effectively users can understand and apply ML results for some practical purpose}. However, we struggle to create a definition for the usability of the training data. Does a user learn data? Although a user uses data, is some data more usable than others, or is that more a matter of data quality and data appropriateness? 

When processing the abstract and titles for usability, we noted that many authors used usability as more a binary term meaning applicability---e.g., usability means that data can be used to train a model. We disagree with this use of usability, as usability is more appropriate as a user-centered qualitative concept. If we exclude general applicability, we find it hard to define usability of data. 

Other combinations of system elements and NFRs can be defined similarly. 
We can see that all NFRs can be defined over the whole system,  reflecting the scope of NFRs over traditional systems. Almost all apply over the model, and most to the algorithm. Fewer NFRs apply to the training data and the results, but there is no clear pattern here. Some apply to both, others only to one.

We are working towards a framework for the definition of each NFR over each part, including a checklist on which part of the system a particular NFR can be defined. We hope that such exploration can lead to a deeper understanding of each NFR and their application to ML systems.


%% file: Threats.tex
\smallskip\noindent\textbf{External Validity:}
We have only used Scopus, which may mean we miss relevant papers in other databases. However, Scopus is a meta-database that is rich in content on computer science research from multiple publishers.  We searched papers in Scopus up to September 2021, and there may be newer papers that are missed.  Future secondary studies should repeat the search process.

The search string was confined to a small set of terms and keywords, focusing on only a subset of NFRs. We could have searched for alternative terms and stems like "interoper" for interoperability. However, it would be difficult to find equivalent stems for all NFRs (e.g., security) and may have led to an unmanageable increase in the volume of papers without a significant increase in relevant results. Our goal is not to make a conclusive statement on the number of publications, but to gain an approximate idea of the interest in each NFR. A sample is sufficient for such purposes. 

\smallskip\noindent\textbf{Internal Validity:}
There is potential bias in determining paper inclusion. To mitigate this risk, we defined shared inclusion criteria, each of the authors went through each title and abstract separately, and we made a collective decision in cases of disagreement. Our ICR results are often good, and performing a second sample yielded consistent or better ICR scores for all but one NFR.

The clusters we created may be subjective to our experiences and opinions. NFRs could be arranged differently, but we believe our clusters are a good starting point to help organize and direct future research. Further work may add to or adjust the clusters as new evidence is found.

Our consideration of the scope of NFR definitions may also be subjective.  We made these judgements in agreement between all authors, discussing difficult cases. We have tried to justify our selection for a sample of NFRs. Future work will adjust our scoping decisions when more evidence or examples are found.

%% file: Conclusion.tex
In this work, we aimed at understanding and exploring definitions, scope, and the extent of existing research on NFRs for ML systems, as we believe that both the research community and  industry  lack knowledge on NFRs for ML systems compared to understanding in traditional systems. The results show that researchers have focused on many NFRs for ML systems, but the amount of attention directed to each NFR differs drastically. Some NFRs received more attention and were explored more (e.g., peformance, accuracy, efficiency) compared to  other NFRs (e.g., maintainability, traceability). Although such differences were expected, it is useful estimate interest with concrete numbers. 

We created six clusters of NFRs based on the similarity of characteristics and meaning of NFRs, and one cluster of NFRs which does not share similar properties,   with the objective of helping researchers to focus on a particular cluster for their future systematic review studies. These clusters will also help practitioners to understand the similarity of NFRs and provide them a direction on which NFRs they need to consider while developing ML systems. 

We defined NFRs over different granular levels of the ML systems based on the meaning and purpose of those NFRs.  This can help practitioners to understand on which part of the ML system a particular NFR can be considered while developing ML systems. Our future work includes a comprehensive mapping study to identify the current state-of-the-art on selected NFRs for ML systems research, and a framework to guide consideration of NFRs over different elements of ML systems.